\documentclass[aps,prl,reprint,showpacs,superscriptaddress]{revtex4-1} 
\usepackage{graphicx}
\usepackage{amsmath,mathtools,verbatim}
\usepackage{amssymb}
\usepackage{bm} 
\usepackage{epstopdf}
\usepackage{braket}
\usepackage[colorlinks=true, citecolor=blue, urlcolor=black ]{hyperref}
\usepackage{xcolor}
\usepackage{hyperref}
\usepackage{url}
\usepackage[normalem]{ulem}

\newcommand{\hide}[1]{}

\newcommand{\be}{\begin{equation}}
\newcommand{\ee}{\end{equation}}
\newcommand{\bqa}{\begin{eqnarray}}
\newcommand{\eqa}{\end{eqnarray}}

\newcommand{\Imperial}{
Physics Department, Blackett Laboratory, Imperial College London, Prince Consort Road, SW7 2AZ, United Kingdom}
\newcommand{\IITB}{Department of Physics, Indian Institute of Technology Bombay, Powai, Mumbai 400076, India}
\newcommand{\HZ}{Helmholtz--Zentrum Dresden-Rossendorf, Bautzner Landstra{\ss}e 400, 01328 Dresden, Germany}

\begin{document}

\title{Breakdown of Temporal Coherence in Photon Condensates}

\author{Yijun Tang} \affiliation{\Imperial}
\author{Himadri S. Dhar}\affiliation{\IITB} 
\author{Rupert F. Oulton}\affiliation{\Imperial}
\author{Robert A. Nyman}\affiliation{\Imperial} \author{Florian Mintert}\affiliation{\Imperial}\affiliation{\HZ}

\date{\today}    

\begin{abstract}
The temporal coherence of an ideal Bose gas increases as the system approaches the Bose-Einstein condensation threshold from below,
with coherence time diverging at the critical point. 
However, counter-examples have been observed for condensates of photons formed in an externally pumped, dye-filled microcavity, wherein the coherence time decreases 
%\sout{drastically}
{rapidly}
for increasing particle number above threshold.
This paper establishes intermode correlations as the central explanation for the experimentally observed dramatic decrease in the coherence time beyond critical pump power.

\end{abstract}     
\maketitle

Coherence is an ubiquitous feature of wave mechanics that plays an important role in  various fields of science, ranging from quantum computing~\cite{DiVincenzo1999} to neurobiology~\cite{Rodriguez1999}.
Originally the notion of coherence had mostly been
discussed in experiments involving
light~\cite{Wolf1995}, but since the development of quantum theory, spatial and temporal coherence can also be observed in massive particles with particularly pronounced wave-like behaviour, such as Bose-Einstein condensates (BEC)~\cite{Ketterle2002}.

While lasers are the best known sources of coherent light~\cite{Wolf1995}, unlike matter waves or BECs, they primarily operate deep in the nonequilibrium regime. However, in recent years, photonic systems such as polaritons and photons in a microcavity and excitons in semiconductors, have been curated to undergo a 
%\sout{non-equilibrium} 
phase transition that spontaneously creates a spatio-temporal coherent light similar to an atomic BEC~\cite{Kasprzak2006,Klaers2010b,Marelic2015,Hakala2018,Barland2021}.
While still driven-dissipative in nature, in contrast to lasers, 
these systems operate in a quasi-equilibrium regime and offer rich dynamics and coherent phenomena~\cite{Bloch2022}, ranging from simultaneous condensation in multiple modes~\cite{Walker2018},
{vortex-like structures~\cite{Dhar2021}}, critical behaviour~\cite{Walker2019} and 
particle correlation~\cite{Schmitt2014,Walker2020} in photons to superfluidity~\cite{Sanvitto2010} and topological lasing~\cite{Amelio2020} in polaritons.

Our work here focuses on the temporal coherence properties of Bose-Einstein condensates of photons, formed inside a dye-filled microcavity~\cite{Klaers2010b,Marelic2015}. 
A defining characteristic of photon BECs is that thermalization is achieved not due to some intrinsic nonlinearity and particle interaction but rather through incoherent interactions between the photon gas and the dye molecules, driven by an incoherent pump~\cite{Klaers2010,Nyman2014}. 
Phase coherence in these condensates result{s} from photon emission from molecules~\cite{Snoke2013}, and experimental observations show a
spontaneous transition from short-range coherence at thermal scale to long-range coherence in the BEC phase~\cite{Marelic2016,Damm2017}.

The temporal coherence time $\tau$ of light inside the cavity is experimentally observed to grow with increasing photon number $n$ as the system approaches the BEC threshold from below~\cite{Marelic2016,Damm2017}.
This is consistent with the well-known Schawlow–Townes linewidth $\tau \propto n$ for $n \gg 1$~\cite{Schawlow1959}.
A microscopic model of photon BEC in a single mode cavity, gives a closed theoretical expression for $\tau$, which indeed reduces to the Schawlow–Townes 
limit for large $n$~\cite{Kirton2015}. 
% %%
However, experimental observations show that this is no longer the case for high photon numbers beyond the transition threshold~\cite{Marelic2016, Walker2018}. In contrast, the temporal coherence breaks down and the coherence time  decreases dramatically as the pump power is increased above the BEC critical point.

Based on a concurrently developed framework for photon-photon correlations in driven-dissipative photon gases~\cite{Tang2023}, 
we show that intermode correlations and competition of different photon modes for access to molecular excitations are the primary reason for the breakdown in 
% \hil{the growth of} \flo{(drop)} 
temporal coherence in photon condensates formed inside a dye-filled microcavity.

A small decrease of molecular excitations can have a substantial impact on the photonic state inside the cavity, such as loss of population or decondensation of individual modes~\cite{Hesten2018}.
Intermode correlations, even if weak, can alter the molecular excitation profile, and 
%we will show here that
we show that this can cause a loss of temporal coherence in the condensed mode without affecting the mode population.
This allows the coherence time $\tau$ to decrease
even when the mode population $n$ is large, 
%thus violating 
contradicting the Schawlow–Townes 
%\sout{\flo{law}}
{limit},
as observed in experimental studies of temporal coherence in photonic BECs formed inside a dye-filled microcavity~\cite{Walker2018}.
Moreover, this observation is larger than any loss of coherence that may arise from small nonlinear effects such as the Henry factor~\cite{Henry1982} or thermo-optical effects~\cite{Klaers2011,Alaeian2017}.

The properties of photon condensates are dependent on a few key parameters, the foremost being the cutoff frequency $\omega_0$ of the microcavity, which then sets the range of available photon modes in the cavity, with frequencies $\omega_p$. The properties of the dye filling the cavity govern the rate of absorption $\mathcal{A}_p$ and emission $\mathcal{E}_p$ for the different photon modes.
The dye is incoherently excited with a pump rate $\Gamma_\uparrow$, which introduces more photons inside the cavity via spontaneous and stimulated processes, while photons and molecules are simultaneously lost with rates $\kappa$ and $\Gamma_\downarrow$, respectively. 
The temporal coherence between a pair of modes $p$ and $q$ of a photonic system can be characterized in terms of the two-time correlation function
\begin{equation}
c_{pq}(t_2-t_1)=\langle\hat{a}_p^\dagger(t_2)\hat{a}_{q}(t_1)\rangle\ ,
 \end{equation}
where $\hat{a}_k$($\hat{a}_k^\dag$) are the annihilation (creation) operator for the $k^{th}$ photon mode,
which, in steady-state experiments, depends only on the time-difference $t_2-t_1$.
To represent correlation functions for multiple modes in a compact form, we can consider
%a vector
{the vectors $\textbf{c}_p$ with elements 
$[\textbf{c}_p]_q = c_{pq}e^{-i\omega_pt}$.}
For photon condensates, {each of the vectors} $\textbf{c}_p$ satisfies the equation~\cite{Tang2023}
\be
\label{temporal}
\frac{d{\textbf{c}_p(t)}}{dt}=-
\frac{1}{2}\bigg(\kappa+\mathbf{A}\mathbf{h}-(\mathbf{A}+\mathbf{E})\mathbf{f}\bigg)\textbf{c}_p(t)\ ,
\ee
where $\mathbf{A}$ and $\mathbf{E}$ are diagonal matrices with elements $\mathcal{A}_p$ 
and  $\mathcal{E}_p$, which
are the absorption and emission rate of photon mode $p$, respectively.
The matrices $\mathbf{h}$ and $\mathbf{f}$ have elements
\bqa
\textbf{h}_{pq}&=&\int d\mu(\mathbf{r})\ \psi_{p}^\ast(\mathbf{r})\psi_{q}(\mathbf{r}),\ \mbox{and}
\label{Eq:h}
\\
\textbf{f}_{pq}&=&\int d\mu(\mathbf{r})\ \psi_{p}^\ast(\mathbf{r})\psi_{q}(\mathbf{r})\ f(\mathbf{r}),
\label{Eq:f}
\eqa
which are defined in terms of the Hermite-Gaussian mode functions $\psi_{p}(\mathbf{r})$ of the cavity, the molecular density $\mu(\mathbf{r})$ and the local excitation fraction $f(\mathbf{r})$ of the molecules, at position $\textbf{r}$ in the cavity plane.
While $\psi_{p}(\mathbf{r})$ and $\mu(\mathbf{r})$ remain unchanged for a specific experimental setting, $f(\mathbf{r})$ is dependent on the external pump power and other system parameters. As such, while the parameters $\kappa$, $\mathbf{E}$, $\mathbf{A}$ and $\mathbf{h}$ are constant, the excitation matrix $\mathbf{f}$ is time dependent.
However, for most experiments, the long-time, steady state of the system is often of primary interest,
and this state is the initial state in the following discussion.
Temporal changes in the excitation of the dye molecules 
% \hil{at $t$} \flo{(drop)} 
are thus negligible.

Eq.~\eqref{temporal} indicates that in the absence of any dye inside the cavity the temporal coherence of each individual mode would simply decay on a timescale proportional to the photon loss rate $\kappa$. The presence of the dye inside the cavity and the ensuing absorption and emission of photons by the dye molecules give rise to additional terms.
In particular, irrespective of the photonic state in the cavity, the absorption of photons by the molecules leads to an additional decay due to the term $\mathbf{A}\mathbf{h}$, which accelerates the decay of coherence. On the other hand, the temporal coherence in the system is boosted by the term $(\mathbf{A}+\mathbf{E})\mathbf{f}$, which is intuitively expected as the coherence increases as the molecular excitations increase and the photon gas in the cavity is driven towards the BEC phase transition.  
These general observations apply equally well to the case of single mode systems~\cite{Kirton2015}.

\begin{figure}[t] 
\center
\includegraphics[width=0.49\textwidth]{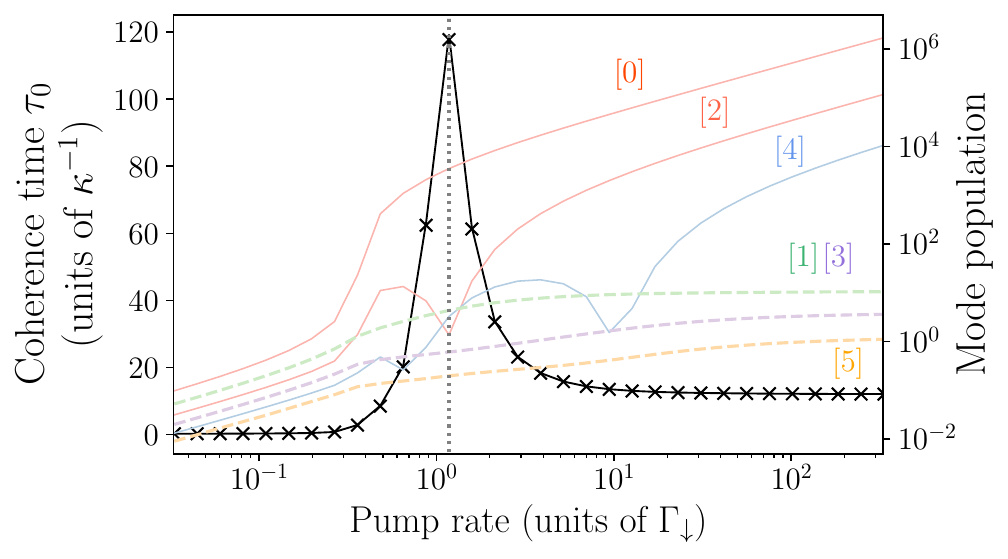} 
\caption{
% \flo{the lines in the plot are very thin.}
The breakdown of temporal coherence in photon condensates. 
The figure shows the ground state coherence time $\tau_0$ as a function of the pump rate $\Gamma_\uparrow$, obtained from
% {via}
%by solving 
Eq.~\eqref{temporal}  (solid-black line) and {via}
%
%%
%the qualitative description in 
Eq.~\eqref{eq:tau} (black-crosses). 
In the background, the steady state population of the lowest six photon modes $([0,5])$ inside the cavity for different pump rates are shown. The coherence of the condensed ground state decreases at the condensation threshold of mode ([2]) (dotted-gray line).
For the numerical study, the cavity cutoff (the ground state) frequency is $\omega_0 = 535$ THz, with mode spacing of $\Delta\omega = 1.7$ THz. The photon loss and the molecule decay rate are equal to $\kappa \approx 0.2$ THz and $\Gamma_\downarrow \approx 3 \times 10^{-5}$ THz, respectively. The absorption and emission rates are calculated from experimental data~\cite{Nyman2017}. }
\label{temp_coh_plot}
\end{figure}

The sudden loss of temporal coherence can be seen
 in Fig.~\ref{temp_coh_plot},
that depicts the coherence time obtained from Eq.~\eqref{temporal}.
The different mode populations are given by the steady state solution of the master equation for a mode-coupled, 
%\sout{nonequilibrium} 
{microscopic} model of photon condensation~\cite{Keeling2016,Dhar2021}.
The figure shows the ground state coherence time $\tau_0$ as a function of the pump rate $\Gamma_\uparrow$, both below and above the BEC threshold.
The condensation of the ground state (mode [0]) leads to a sharp growth of the coherence time. 
Importantly, the coherence reaches it maximum around $1.18~\Gamma_\downarrow$, at the point when the population of the second excited state (mode [2]) starts growing.
Subsequent increase in $\Gamma_\uparrow$ leads to a sharp decline in the temporal coherence of the ground state and a decrease of $\tau_0$, 
consistent with recent experimental observations~\cite{Walker2018}.

A simple qualitative picture for the coherence time emerges if the matrices $\mathbf{h}$ and $\mathbf{f}$
can be considered to be diagonal.
This approximation works well, since Eqs.~\eqref{Eq:h} and \eqref{Eq:f} involve an integration over the density of dye molecules;
for off-diagonal elements, the integrand is oscillatory, whereas it is non-negative for diagonal elements.
The integration thus results in a cancellation of terms with different phase for off-diagonal elements
in contrast to the accumulation of non-negative contributions for the diagonal terms~\cite{comment}. 
Therefore, the solutions of Eq.~\eqref{temporal} are expected to behave qualitatively similar to the single-mode case,
such that the two-time correlation function $\textbf{c}_p(t)$ for mode $p$ 
(with $p=0$ for the ground state) will follow an exponential behavior, with the coherence time given by
\be
\tau_{p}=\frac{2}{\left(\kappa+\mathcal{A}_{p}\mathbf{h}_{pp}-(\mathcal{A}_{p}+\mathcal{E}_{p})\mathbf{f}_{pp}\right)}\ .
\label{eq:tau}
\ee
It is instructive to compare this to the  steady state population
\begin{eqnarray}
n_0 = c_{00}(0) &=& \frac{\mathcal{E}_{0}\mathbf{f}_{00}+\sum_{j\neq 0}(\mathcal{E}_j+\mathcal{A}_j)\textbf{f}_{0j}n_{j0}}{\kappa+\mathcal{A}_0 \textbf{h}_{00} - (\mathcal{A}_0 + \mathcal{E}_0)\textbf{f}_{00}},
\label{eq:n0}
\end{eqnarray}
of the ground state,
where $n_{j0} = c_{j0}(0)$, is the intermode photon correlation.
Since $\tau_{p}$ in Eq.~\eqref{eq:tau} and $n_0 $ in Eq.~\eqref{eq:n0} have the same denominator, one would expect qualitatively similar behavior for the coherence time and the ground state population.

Far below the condensation threshold the molecular excitation at steady state is small. However, as the molecular excitation 
%\sout{fraction} 
$\textbf{f}_{00}$ increases with growing pump power, the term $(\mathcal{A}_{0}+\mathcal{E}_{0})\mathbf{f}_{00}$ becomes larger, thus lowering the denominator in Eqs.~\eqref{eq:tau} and \eqref{eq:n0}.
This leads to increase of both the photon population $n_0$ and the coherence time $\tau_0$ of the ground state.
The system transitions to the Bose-Einstein condensate phase as the denominator approaches zero {i.e., $\mathbf{f}_{00}$ is clamped to the value $(\kappa+\mathcal{A}_{p}\mathbf{h}_{pp})/(\mathcal{A}_{0}+\mathcal{E}_{0})$,} and both $n_0$ and $\tau_0$ diverge.
However, if $\mathbf{f}_{00}$ is unclamped due to mode competition and decreases even slightly, the denominator in Eqs.~\eqref{eq:tau} and \eqref{eq:n0} increases, leading to a loss of coherence time $\tau_0$.

% \flo{Fig.~\ref{eq:tau} shows the clamping and the subsequent unclamping of the molecular excitation in terms of the contributions $\kappa+\mathcal{A}_{p}\mathbf{h}_{pp}$ and $(\mathcal{A}_{p}+\mathcal{E}_{p})\mathbf{f}_{pp}$ to the coherence time in Eq.~\eqref
% {eq:tau}.
% (for the axis label we could just say `clamping (units of $\kappa$.)
% }

\begin{figure}[h] 
\center
\includegraphics[width=0.43\textwidth]{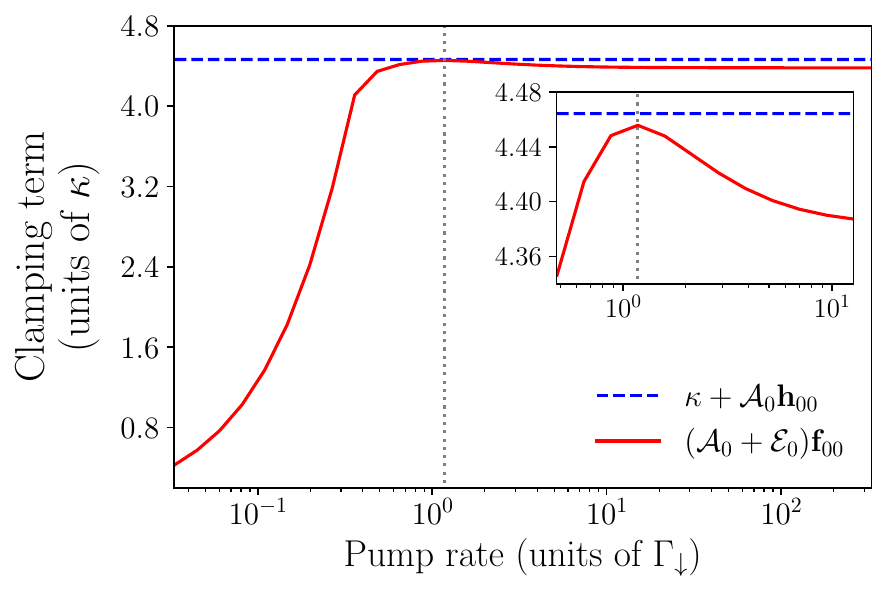} 
\caption{
% \sout{Decrease in effective molecular excitation corresponding to the ground state mode.} 
{Variation of the term $(\mathcal{A}_0 + \mathcal{E}_0)\textbf{f}_{00}$ in Eq.~(\ref{eq:tau}}
% \hsd{Variation of the clamping term $(\mathcal{A}_0 + \mathcal{E}_0)\textbf{f}_{00}$}
%\hil{$\tilde{f}_0$} \flo{(Is this symbol defined?)} 
as a function of the pump rate $\Gamma_\uparrow$ (solid-red line), as it converges to
{$(\kappa+\mathbf{A}_{0})\mathbf{h}_{0}$}
(dashed-blue line). 
%$\tilde{f}_0$ 
The term starts decreasing at the condensation threshold of mode ([2]) (dotted-gray line).
Importantly, this decrease is not accompanied by drop in ground state population $n_0$, as evident from Fig.~\ref{temp_coh_plot}. 
The inset zooms in on the drop of the term around the threshold. 
% \hsd{The y-axis in both the plots are in units of $\kappa$.}
%%
The system parameters are the same as Fig.~\ref{temp_coh_plot}.
}
\label{mol_excite}
\end{figure}

{Figure~\ref{mol_excite} shows the clamping and unclamping of the molecular excitation $\textbf{f}_{00}$ through the variation of the term $(\mathcal{A}_{0}+\mathcal{E}_{0})\mathbf{f}_{00}$ with increasing pump power. }
%
% The clamping and the subsequent unclamping of the molecular excitation %%\sout{fraction} 
% $\textbf{f}_{00}$ is shown in Fig.~\ref{mol_excite}. The figure shows the change of the term $(\mathcal{A}_0+\mathcal{E}_0)\textbf{f}_{00}$ with increasing pump power. 
% 
As the ground state (mode [0]) condenses, the term rapidly approaches its threshold value $(\kappa+\mathcal{A}_{0})\mathbf{h}_{00}$.
Comparison with Fig.~\ref{temp_coh_plot} shows that condensation occurs when $\textbf{f}_{00}$ is clamped to a value close to $(\kappa+\mathcal{A}_{0})\mathbf{h}_{00}/(\mathcal{A}_{0}+\mathcal{E}_{0})$.
Notably, Eq.~(\ref{eq:tau}) predicts that the coherence time $\tau_0$ rises sharply in this region. 
In a single-mode system, there is no mechanism that would break this clamping, but unclamping can occur in a multimode system.
From Fig.~\ref{mol_excite}, close to the value of $1.18~\Gamma_\downarrow$, the excitation profile changes suddenly, {\it i.e.} $\textbf{f}_{00}$  is unclamped due to mode-competition arising from the increase in population of the second-excited state (mode [2]), which leads to a decrease
of the coherence time $\tau_0$.

An important question here is why does the ground state population $n_0$ not decrease along with the coherence time $\tau_0$, when the molecular excitation 
%\sout{fraction} 
$\textbf{f}_{00}$ drops below the threshold. The answer lies in the intermode correlations, which acts to preserve the ground state population,
as shown in Eq.~\eqref{eq:n0}.
In the absence of correlations between the modes (i.e. for $n_{j0} = 0$),  the ground state population is given by the contribution $n_e=n_0|_{n_{j0} = 0}$, 
which is driven purely by stimulated emission term $\mathcal{E}_{0}\mathbf{f}_{00}$.
As such, a drop in the molecular excitation $\mathbf{f}_{00}$ would not only result in a decrease of $\mathcal{E}_{0}\mathbf{f}_{00}$, but also in an increase of the denominator, $\kappa+\mathcal{A}_0 \textbf{h}_{00} - (\mathcal{A}_0 + \mathcal{E}_0)\textbf{f}_{00}$.
Hence, for $n_{j0} = 0$ the ground state population $n_{e}$ decreases along with the coherence time $\tau_0$, whenever $\mathbf{f}_{00}$ is unclamped.

The situation is drastically different in the presence of intermode correlations ($n_{j0} \neq 0$), where
any drop in steady state population $n_0$ due to a decrease in the contribution $n_e$ is countered by an {increase in the contribution of $n_c=n_0|_{n_{e} = 0}$.} 
Importantly, 
$n_c$ is proportional to the 
%mode correlation-dependent 
{mode-correlation dependent}
term $\sum_{j\neq 0}(\mathcal{E}_j+\mathcal{A}_j)\textbf{f}_{0j}n_{j0}$ in Eq.~(\ref{eq:n0}).
Hence, even though the competition for excitation between the modes leads to unclamping of  $\textbf{f}_{00}$ and a decrease in $n_{e}$,
the correlations between the modes contribute to $n_c$ and compensate 
% \flo{compensate (the correlations)} 
for the loss in ground state photons. 
\begin{figure*}[t] 
\center
\includegraphics[width=0.7\textwidth]{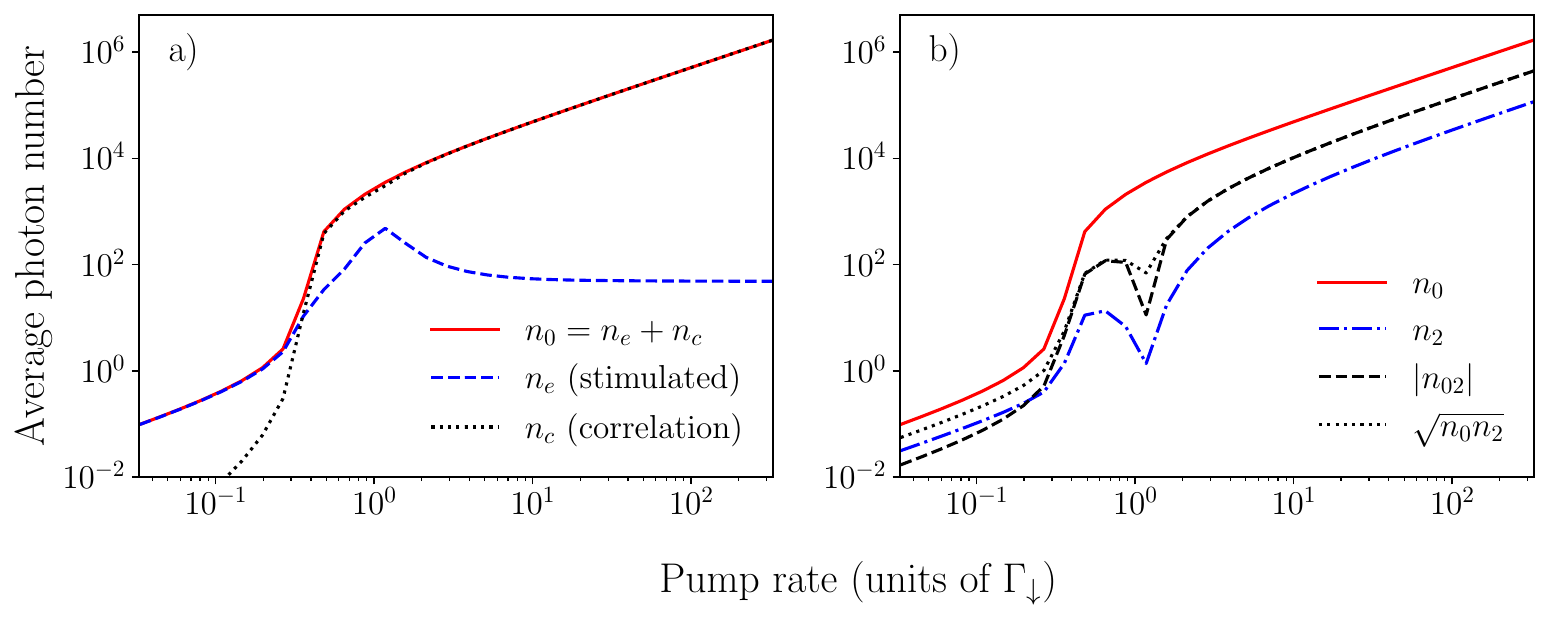} 
\caption{
The contributing terms in the ground state population, the second excited mode and the intermode correlation {as  a function of rate of pumping}. 
a) 
Ground state population, $n_0$ = $n_e$ + $n_c$ (red-solid), along with the terms arising due to stimulated emission $n_{e} \propto  \mathcal{E}_0 \textbf{f}_{00}$ (blue-dashed) and intermode correlation $n_{c} \propto \sum_{j\neq 0} (\mathcal{E}_j+\mathcal{A}_j)\textbf{f}_{0j}n_{j0}$ (black-dotted).
b) 
Population of the ground state $n_{0}$  (red-solid) and 
the second-excited mode $n_{2}$ (blue-dashed-dots), along with the absolute value of the mode correlation $|n_{02}|$ (black-dashed) and  $\sqrt{n_0 n_2}$ (black-dotted). The system parameters are the same as in Fig.~\ref{temp_coh_plot}.}
\label{photon_pop}
\end{figure*}

Fig.~\ref{photon_pop} shows the behavior of the key terms in the expression for the ground state photon population $n_{0}$ (Eq.~(\ref{eq:n0}).
The two main contributing terms, the correlation independent term $n_{e}$
and the correlation dependent $n_c$,
are shown in Fig.~\ref{photon_pop}(a).
{The contribution from $n_c$ is substantially larger than that of the correlation independent term $n_{e}$.}
As such, even though the term $n_{e}$ decreases when ${\textbf{f}}_{00}$ is unclamped, the condensed ground state mode population $n_{0}$ remains unaffected. However, the condensed mode is no longer temporally coherent as $\tau_0$ decreases with drop in ${\textbf{f}}_{00}$.
Furthermore, Fig.~\ref{photon_pop}(b), also compares the ground state population $n_{0}$ with the second mode population $n_{2}$ and the intermode coherence $n_{20}$.
A key takeaway here is that the coherence term $n_{20}$ is large, where $|n_{20}|$ is equal to $\sqrt{n_0n_2}$ for high pump powers, which implies perfect phase coherence between the modes.
Therefore, even though the off-diagonal 
%\sout{population fraction} 
excitation matrix element $\textbf{f}_{02}$ is more than an order of magnitude smaller than the diagonal term $\textbf{f}_{00}$ 
(with $|\textbf{f}_{02}/\textbf{f}_{00}| \approx 10^{-2}-10^{-4}$), 
the large value of $n_{20}$ ensures that the {correlation} dependent term $n_c$ in Eq.~(\ref{eq:n0}) is large enough to ensure that the ground state photon number $n_{0}$ remains unchanged. 

Another question is whether the condensing mode is a superposition of several modes.
Diagonalizing the correlation matrix $n_{ij} = c_{ij}(0)$ shows that the proportion of the photons in the ground state $p_0$ is rather high ($p_0 \ge 0.93$), compared to modes [2] ($p_2 \le 0.064$) and mode [4] ($p_4 \le 6\times10^{-3}$), and the average population of the odd modes are negligible.
As such, an immediate effect of the intermode correlation is 
%\sout{the fact} 
that for a pump focused at the center of the cavity only even modes ([0],[2],[4]) condense. The wavefunctions of even modes peak at the center and therefore have greater access to excited molecules and also correlate strongly with the ground state mode, as compared to the odd modes ([1],[3],[5]). 

The intermode {correlations} also give rise to other effects. For instance, the population $n_2$ of the second-excited mode increases when the ground state mode condenses, for pump rate values %$0.2\Gamma_\downarrow-0.5\Gamma_\downarrow$
{between $0.2\Gamma_\downarrow$ and $0.5\Gamma_\downarrow$}, as shown in Figs.~\ref{temp_coh_plot} and \ref{photon_pop}(b).
The explanation lies in the intermode correlation between the ground and second excited mode, which is anti-correlated ($n_{20} < 0$) at low pump powers. 
However, in this regime the {off-diagonal} excitation 
%\sout{fraction} \hsd
{term} $\textbf{f}_{20}$ 
% \flo{bold face? why fraction?} 
is also negative, which leads to a net positive contribution to the population, as shown by the absolute value of $n_{20}$ in Fig.~\ref{photon_pop}(b). This rise in $n_2$ is purely driven by the correlations, as the contribution due to stimulated emission $\mathcal{E}_2 \textbf{f}_{22}$ remains significantly smaller. 
Another interesting observation is that the temporal coherence does not completely disappear at higher pump powers in Fig.~\ref{temp_coh_plot}.
This is due to the fact that the number of photons arising from stimulated emission $n_e$ does not vanish but converges to a significant steady state value at high pump powers, as shown in Fig.~\ref{photon_pop}.

The presently used non-equilibrium model for photon condensation~\cite{Tang2023} provides the footing to identify intermode correlations as the cause for changes in molecular excitation profile, 
which ultimately leads to the breakdown of temporal coherence, 
% in violation of
thus contradicting the behaviour predicted by
the Schawlow-Townes limit.
%\flo{law}. % limit. %% law is typically not used with Schawlow_Townes }
The mechanism studied here is quite general and %such
{similar} effects of intermode correlations could be engineered in other optical systems, where it could lead to generation of partially coherent light, which is a powerful resource in imaging~\cite{Singh2015} and communications~\cite{Huang2021}.
A future direction is to investigate temporal coherence during other nonequilibrium phenomena such as decondensation, where a higher mode forces a lower condensed mode to lose its population. In such a regime, the simultaneous loss of temporal coherence and decondensation in a pair of higher energy modes, could lead to favorable changes in the molecular excitation profile and recoherence in the ground state mode. Other avenues include the study of quantum correlations developing between coupled, but spatially separated modes~\cite{Kurtscheid2019}, especially in condensates with only few photons. These will prove useful in manipulation of quantum states of light~\cite{Vretenar2021}, with potential application in interferometry and metrology. 

The authors acknowledge financial support from the European Commission via the PhoQuS project (H2020-FETFLAG-2018-03) number 820392 and the EPSRC (UK) through the grants EP/S000755/1. HSD acknowledges financial support from SERB-DST, India via a Core Research Grant CRG/2021/008918 and the Industrial Research \& Consultancy Centre, IIT Bombay via grant (RD/0521-IRCCSH0-001) number 2021289. 
% \textbf{(Add other grants.)}


\begin{thebibliography}{99}
\bibitem{comment} The approximation made while solving Eq.~(\ref{temporal}) has been verified numerically, where the contributions of the off-diagonal terms in estimating the coherence time is found to be negligible compared to the diagonal terms.
\end{thebibliography}
\end{document}